\documentclass[aip,rsi,reprint,graphicx]{revtex4-2} % for checking your page length; this is only an ESTIMATE!
\usepackage{graphicx}% Include figure files
\usepackage{dcolumn}% Align table columns on decimal point
\usepackage{bm}% bold math
\usepackage{amsmath}

\usepackage[utf8]{inputenc}
\usepackage[T1]{fontenc}
\usepackage{mathptmx}
\usepackage{etoolbox}

\usepackage{tikz} 
\usepackage{tkz-euclide}
\usepackage{tabularx}
\usepackage{pgfplots}
\pgfplotsset{compat=1.7}
\usepackage{subcaption}
\usepgfplotslibrary{colormaps}
\pgfplotsset{
	colormap={mycolormap}{color=(white) color=(lightgray) color=(white)},
	colormap={whites}{color=(white) color=(white) color=(white)},
}

\makeatletter
\def\@email#1#2{%
 \endgroup
 \patchcmd{\titleblock@produce}
  {\frontmatter@RRAPformat}
  {\frontmatter@RRAPformat{\produce@RRAP{*#1\href{mailto:#2}{#2}}}\frontmatter@RRAPformat}
  {}{}
}%
\makeatother

\newcommand{\rmd}[0]{\textrm{d}}
\newcommand{\rmi}[0]{\textrm{i}}
\begin{document}

\preprint{AIP/123-QED}

\title[Validating and optimising mismatch tolerance of Doppler backscattering measurements with the beam model]{Validating and optimising mismatch tolerance of Doppler backscattering measurements with the beam model}
% Force line breaks with \\

\author{V.H. Hall-Chen}
\affiliation{ 
Institute of High Performance Computing, Singapore 138632, Singapore
}
\affiliation{%
UKAEA/CCFE, Culham Science Centre, Abingdon, Oxon, OX14 3DB, UK
}%
\affiliation{ 
Rudolf Peierls Centre for Theoretical Physics, University of Oxford, Oxford OX1 3PU, UK
}
\email[The author to whom correspondence may be addressed: ]{valerian\_hall-chen@ihpc.a-star.edu.sg.}


\author{J. Damba}
\affiliation{%
Department of Physics and Astronomy, University of California, Los Angeles, CA 90095, USA
}%

\author{F.I. Parra}%
\affiliation{ 
Princeton Plasma Physics Laboratory, Princeton, NJ 08540, USA%\\This line break forced with \textbackslash\textbackslash
}%

\author{Q.T. Pratt}
\affiliation{%
Department of Physics and Astronomy, University of California, Los Angeles, CA 90095, USA
}%

\author{C.A. Michael} 
\affiliation{% https://orcid.org/0000-0003-1804-870X
	Department of Physics and Astronomy, University of California, Los Angeles, CA 90095, USA
}%

\author{S. Peng}
\affiliation{%
	UKAEA/CCFE, Culham Science Centre, Abingdon, Oxon, OX14 3DB, UK
}%

\author{T.L. Rhodes}
\affiliation{%
Department of Physics and Astronomy, University of California, Los Angeles, CA 90095, USA
}%

%%%%%

\author{N.A. Crocker}
\affiliation{%
Department of Physics and Astronomy, University of California, Los Angeles, CA 90095, USA
}%

\author{J.C. Hillesheim}
\affiliation{%
UKAEA/CCFE, Culham Science Centre, Abingdon, Oxon, OX14 3DB, UK
}%

\author{R. Hong}
\affiliation{%
Department of Physics and Astronomy, University of California, Los Angeles, CA 90095, USA
}%

\author{S. Ni} % https://orcid.org/0000-0001-9616-7945
\affiliation{%
	Cavendish Laboratory, University of Cambridge, Cambridge CB3 0HE, UK
}%

\author{W.A. Peebles}
\affiliation{%
Department of Physics and Astronomy, University of California, Los Angeles, CA 90095, USA
}%

\author{C.E. Png}
\affiliation{ % https://orcid.org/0000-0002-7797-1863
Institute of High Performance Computing, Singapore 138632, Singapore
}

\author{J. Ruiz Ruiz}
\affiliation{ 
Rudolf Peierls Centre for Theoretical Physics, University of Oxford, Oxford OX1 3PU, UK
}

\date{\today}% It is always \today, today,

\begin{abstract}
We use the beam model of Doppler backscattering (DBS), which was previously derived from beam tracing and the reciprocity theorem, to shed light on mismatch attenuation. This attenuation of the backscattered signal occurs when the wavevector of the probe beam's electric field is not in the plane perpendicular to the magnetic field. Correcting for this effect is important for determining the amplitude of the actual density fluctuations. Previous preliminary comparisons between the model and Mega-Ampere Spherical Tokamak (MAST) plasmas were promising. In this work, we quantitatively account for this effect on DIII-D, a conventional tokamak. We compare the predicted and measured mismatch attenuation in various DIII-D, MAST, and MAST-U plasmas, showing that the beam model is applicable in a wide variety of situations. Finally, we performed a preliminary parameter sweep and found that the mismatch tolerance can be improved by optimising the probe beam's width and curvature at launch. This is potentially a design consideration for new DBS systems.
\end{abstract}

\maketitle

\section{Introduction}
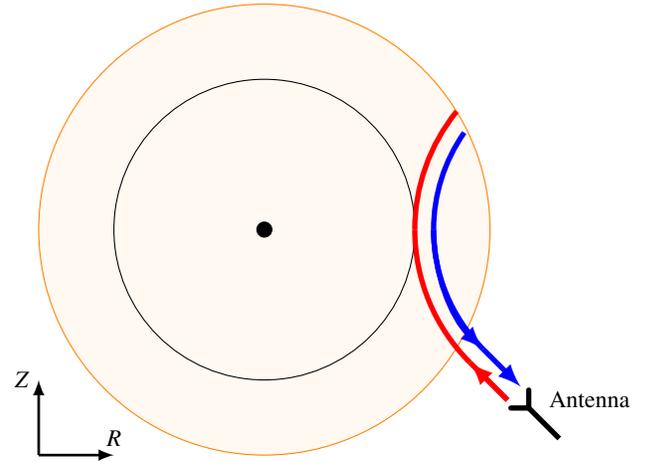
\begin{figure}%{0.45\columnwidth}
\centering
\begin{tikzpicture}
	\filldraw[color=orange,fill=orange!5](0,0) circle (3.0);
	\draw (0,0) circle [black, radius=2];
	\filldraw (0,0) circle [color=black, fill=black, radius=0.1];
	%	\draw[black] (axis cs:4,0) arc (225:135:3cm); %% Very hackily done, need to change the angles to make sure it looks nice for any given viewing angle
	% 	\draw (axis cs:2.121,-2.121) to[out=30,in=210] (axis cs:2.121,2.121);
	
	% Probe beam
	\draw[red, line width = 2pt] (2,0) arc (0:45:-2.5);
	\draw[red, line width = 2pt] (2,0) arc (0:-39:-2.5);
	\draw[red, line width = 2pt] ({2.5-(2.5*cos(45)) +0.5 +2},{-(2.5*sin(45)) -0.5}) -- ({2.5-(2.5*cos(45))+2},{-(2.5*sin(45))});
	\draw[-latex, red, line width = 2pt] ({2.5-(2.5*cos(45)) +0.5 +2},{-(2.5*sin(45)) -0.5}) -- ({2.5-(2.5*cos(45))+2},{-(2.5*sin(45))});
	
	% Backscattered electric field
	\draw[blue, line width = 2pt] (2.25,0) arc (0:-35:-2.25);	
	\draw[blue, line width = 2pt] (2.25,0) arc (0:45:-2.25);
	\draw[-latex,blue, line width = 2pt] (2.25,0) arc (0:45:-2.25);
	\draw[latex-,blue, line width = 2pt] ({2.5-(2.25*cos(45)) +0.5 +2},{-(2.25*sin(45)) -0.5}) -- ({2.5-(2.25*cos(45))+2},{-(2.25*sin(45))});

	% Coordinates
	\draw[-latex, line width = 1pt] (-3,-3) -- (-2,-3) node[above] {$R$};
	\draw[-latex, line width = 1pt] (-3,-3) -- (-3,-2) node[left] {$Z$};
	
	% Antenna
	\draw[{angle 90 reversed}-, black, line width = 2pt] ({2.5-(2.375*cos(45)) +0.55 +2},{-(2.375*sin(45)) -0.55}) -- ({2.5-(2.375*cos(45)) +1.1 +2},{-(2.375*sin(45)) -1.1}) node[midway, above right, rotate=0] {Antenna}; 
\end{tikzpicture}
\caption{Cartoon of Doppler backscattering (DBS) in the poloidal plane of a tokamak. The black dot represents the magnetic axis, the red line the probe beam, and the blue line the backscattered signal.}
\label{fig:DBS_schematic}
\end{figure}

The Doppler backscattering (DBS) diagnostic measures density fluctuations \cite{Hirsch:DBS:2001} and flows \cite{Pratt:DBS:2022} by launching a microwave beam into the plasma and measuring the backscattered power and its Doppler shift; a schematic of this diagnostic is shown in Figure \ref{fig:DBS_schematic}. These density fluctuations typically have spatial scales $1 \lesssim k_{\perp} \rho_i \lesssim 10$, where $k_\perp$ is the turbulence wavenumber perpendicular to the magnetic field and $\rho_i$ is the ion gyroradius. This spatial scale of measured fluctuations scales with electron plasma beta (ratio of thermal and magnetic pressures) \cite{RuizRuiz:RCDR:2022}, $k_{\perp} \rho_i \sim \beta_e^{1 / 2}$. Since $k_\perp \gg k_\parallel$, where $k_\parallel$ is the parallel turbulence wavenumber, we assume that $k_\parallel$ does not contribute to the backscattered signal.

The DBS diagnostic is installed on various fusion experiments, such as tokamaks \cite{Hennequin:DBS:2004, Hillesheim:DBS_MAST:2015, Shi:DBS:2016, Rhodes:DBS:2018} and stellarators \cite{Happel:DBS:2009, Tokuzawa:DBS_LHD:2021}. As such, a better understanding of DBS will improve what we can do with existing hardware worldwide. Developing this understanding is especially challenging as the backscattered power has a complicated dependence on properties of the probe beam and plasma equilibrium \cite{Gusakov:scattering_slab:2004, Bulanin:spatial_spectral_resolution:2006, Hillesheim:DBS_MAST:2015, Hall-Chen:beam_model_DBS:2022}. To develop a quantitative understanding of backscattered signal, we used beam tracing \cite{Pereverzev:Beam_Tracing:1996, Poli:Torbeam:2001} and the reciprocity theorem \cite{Piliya:reciprocity:2002, Gusakov:scattering_slab:2004} to derive the beam model of DBS \cite{Hall-Chen:beam_model_DBS:2022}. This is a theoretical framework for DBS in general geometry, giving insights on the various contributions to the backscattered signal. The model, together with the requisite beam tracer, is implemented in our simulation code, Scotty. 

In this paper, we are concerned with one particular contribution to the backscattered signal: the attenuation of the backscattered signal when the wavevector of the probe beam's electric field is not in the plane perpendicular to the magnetic field. This is quantified by the mismatch angle \cite{Hillesheim:DBS_MAST:2015}, the angle between the probe beam's wavevector and the plane perpendicular to the magnetic field, given by
\begin{equation}
	\sin \theta_{m} = \frac{K_\parallel}{K},
\end{equation}
where $\theta_{m}$ is the mismatch angle, $K_\parallel = \mathbf{K} \cdot \hat{\mathbf{b}}$ is the parallel wavenumber of probe beam, $\hat{\mathbf{b}} = \mathbf{B} / B$ is the unit vector of the magnetic field, $\mathbf{B}$ is the local magnetic field, and $\mathbf{K}$ is the beam wavevector; the magnitudes of the latter two are $B$ and $K$, respectively. Note that the mismatch angle is defined at every point along the central ray of the probe beam's trajectory; the evolution of mismatch angle along the central ray was discussed in previous work \cite{Hillesheim:DBS_MAST:2015}. 

To simplify analysis in this paper, we consider the location where most of the measured backscattering occurs and evaluate the mismatch angle at that point only. It is traditionally assumed that this location is at or near the cut-off \cite{Hirsch:DBS:2001}. Note that the general beam model of DBS \cite{Hall-Chen:beam_model_DBS:2022} does not need to make this assumption. In fact, we showed that in cases where the mismatch angle is zero at one point along the ray's trajectory and large elsewhere, then the backscattered signal is localised to that point, regardless of whether that point is near the cut-off. On the other hand, when the mismatch angle is small along the probe beam's trajectory, we found that various contributions weigh the measured backscattered signal most strongly when the probe beam's wavenumber is minimised \cite{Hall-Chen:beam_model_DBS:2022}, which turns out to be at the cut-off. The shots studied in this paper fall largely into the latter category. Hence, for simplicity, we assume in this paper that the backscattered signal is localised to the point where the wavenumber is minimised, which we nominally refer to as the cut-off location.

\begin{figure*}
	\centering
	\begin{subfigure}{0.45\textwidth}
		\centering
		\begin{tikzpicture}
			\begin{axis}[
				axis equal image,
				hide axis,
				z buffer = sort,
				view = {0}{20},
				scale = 1.5,
				xmin = -4.1,
				xmax = 4.1,
				ymin = -4.1,
				ymax = 3.1,
				zmin = -3.1,
				zmax = 3.1,
				]
				\addplot3[
				surf,
				shader = faceted interp,
				samples = 17,
				samples y = 33,
				domain = -2*pi:0,
				domain y = -2*pi:0,
				colormap name = mycolormap,
				thin,
				](
				{(3+sin(deg(\x)))*cos(deg(\y))},
				{(3+sin(deg(\x)))*sin(deg(\y))},
				{cos(deg(\x))}
				);
				%		\draw[-latex] (axis cs:0,0,0) -- (axis cs:1.8,0,0) node [above, xshift=-1.2cm, yshift=-0.05cm] {\footnotesize Radial};
				%		\draw[-latex] (axis cs:3.75,0,-1.5) to[out=55,in=-55] (axis cs:3.75,0,1.5) node [xshift = 0.6cm, yshift = -1.2cm, rotate = -90] {\footnotesize Poloidal};
				%		\draw[-latex] (axis cs:-2,-5,0) to[out=-10,in=190] (axis cs:2,-5,0) node [xshift = -1.9cm, yshift = -0.4cm] {\footnotesize Toroidal};
				\draw[-latex, blue, ultra thick] (axis cs:-1.0,-4,-0.667) -- (axis cs:1.0,-4,0.667) node [circle, inner sep = 0.2pt, fill=white, above] {\footnotesize $\mathbf{B}$};
				\draw[-latex, red, ultra thick] (axis cs:0,-4,-2) -- (axis cs:0,-4,0) node [circle, inner sep = 0.2pt, fill=white, xshift = -0.3cm, yshift = -0.4cm] {\footnotesize $\mathbf{K}$};		
				\draw[dashed] (axis cs:1.109,-4,-1.664) -- (axis cs:0,-4,0);		
				\draw[black] (axis cs:0,-4,-1) arc (-90:-35:0.5cm) node [circle, inner sep = 0.2pt, fill=white, xshift = -0.05cm, yshift = -0.4cm] {\footnotesize $\theta_m$}; %% Very hackily done, need to change the angles to make sure it looks nice for any given viewing angle
			\end{axis}
		\end{tikzpicture}
	\end{subfigure} 
	\quad
	\begin{subfigure}{0.45\textwidth}
		\centering
		\begin{tikzpicture}
			\begin{axis}[
				axis equal image,
				hide axis,
				z buffer = sort,
				view = {0}{20},
				scale = 1.5,
				xmin = -4.1,
				xmax = 4.1,
				ymin = -4.1,
				ymax = 3.1,
				zmin = -3.1,
				zmax = 3.1,
				]
				\addplot3[
				surf,
				shader = faceted interp,
				samples = 17,
				samples y = 33,
				domain = -2*pi:0,
				domain y = -2*pi:0,
				colormap name = mycolormap,
				thin,
				](
				{(3+sin(deg(\x)))*cos(deg(\y))},
				{(3+sin(deg(\x)))*sin(deg(\y))},
				{cos(deg(\x))}
				);
				%		\draw[-latex] (axis cs:0,0,0) -- (axis cs:1.8,0,0) node [above, xshift=-1.2cm, yshift=-0.05cm] {\footnotesize Radial};
				%		\draw[-latex] (axis cs:3.75,0,-1.5) to[out=55,in=-55] (axis cs:3.75,0,1.5) node [xshift = 0.6cm, yshift = -1.2cm, rotate = -90] {\footnotesize Poloidal};
				%		\draw[-latex] (axis cs:-2,-5,0) to[out=-10,in=190] (axis cs:2,-5,0) node [xshift = -1.9cm, yshift = -0.4cm] {\footnotesize Toroidal};
				\draw[-latex, blue, ultra thick] (axis cs:-1.0,-4,-0.667) -- (axis cs:1.0,-4,0.667) node [circle, inner sep = 0.2pt, fill=white, above] {\footnotesize $\mathbf{B}$};
%				\draw[-latex, red, ultra thick, opacity = 0.2] (axis cs:0,-4,-2) -- (axis cs:0,-4,0);		
				\draw[-latex, red, ultra thick] (axis cs:1.109,-4,-1.664) -- (axis cs:0,-4,0) node [circle, inner sep = 0.2pt, fill=white, xshift = 0.45cm, yshift = -0.1cm] {\footnotesize $\mathbf{K}$};		
				%		\draw[black] (axis cs:0,-4,-1) arc (-90:-35:0.5cm) node [circle, inner sep = 0.2pt, fill=white, xshift = -0.05cm, yshift = -0.4cm] {\footnotesize $\theta_m$}; %% Very hackily done, need to change the angles to make sure it looks nice for any given viewing angle
			\end{axis}
		\end{tikzpicture}	
	\end{subfigure}
	\caption{Schematic to illustrate mismatch. Launching the probe beam with no toroidal propagation leads to a finite mismatch at the cut-off due to the magnetic field's pitch angle (left). With well-chosen toroidal steering (right), the probe beam reaches the cut-off and backscatters off a similar turbulent $\mathbf{k}_\perp$ as before, but with no mismatch.}
	\label{fig:toroidal_steering}
\end{figure*}
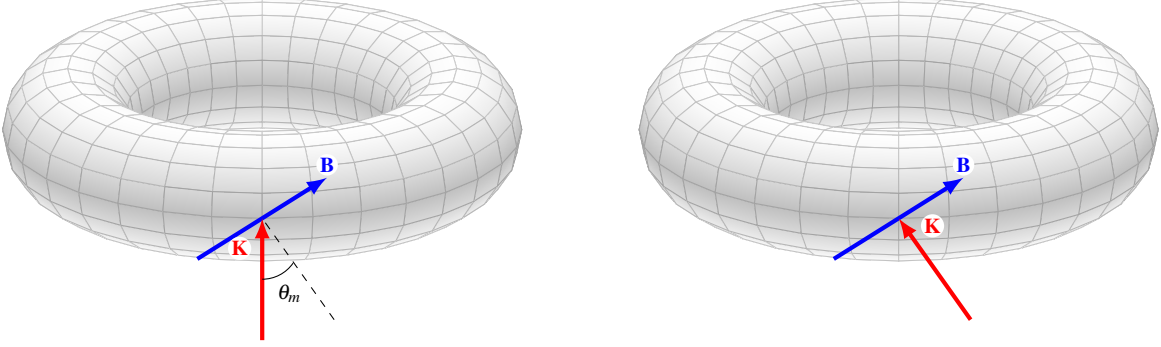

Launching the probe beam entirely in the poloidal plane, with no toroidal propagation, leads to a finite mismatch at the cut-off due to the magnetic field's pitch angle. When the mismatch angle is non-zero, the backscattered signal is reduced \cite{Hillesheim:DBS_MAST:2015, Damba:DBS_DIII-D:2022}. We refer to this reduction of the signal as \emph{mismatch attenuation}. However, with well-chosen toroidal steering, the probe beam reaches the cut-off with no mismatch, Figure \ref{fig:toroidal_steering}.

We begin by introducing the quantitative theoretical underpinnings of mismatch attenuation in subsection \ref{subsection:mismatch_theory}, followed by an explanation of how this effect may be quantitatively measured, subsection \ref{subsection:mismatch_experiment}. These techniques are then used to elucidate the effect of mismatch as measured in DIII-D, MAST, and MAST-U, discussed in subsections \ref{subsection:mismatch_DIII-D}, \ref{subsection:mismatch_MAST}, and \ref{subsection:mismatch_MAST-U}, respectively. We then show how the tolerance to mismatch attenuation may be optimised by tuning the width and curvature of the DBS probe beam in section \ref{section:optimisation}. The prospects of using beam tracing and reciprocity to account for mismatch is discussed in section \ref{section:discussion} and finally, the main findings of this work are summarised in section \ref{section:conclusion}.

\section{Evaluating mismatch attenuation}

\subsection{Theoretical model} \label{subsection:mismatch_theory}
\begin{figure}
	\centering
	\begin{tikzpicture}[>=latex]
		\draw[style=help lines] (0,0) (3,2);
		
		\coordinate (vec1) at (90:2.5);
		\coordinate (vec2) at (0:2.5);
		\coordinate (vec3) at (200:2.5);
		\coordinate (vec4) at (110:2.5);
		\coordinate (vec5) at (140:2.5);
		\coordinate (vec6) at (180:2.5);
		
		\draw[->,thick,black] (0,0) -- (vec1) node[left] {$\hat{\mathbf{g}}$};
		\draw[->,thick,black] (0,0) -- (vec2) node[right] {$\hat{\mathbf{x}}$};
		\draw[->,thick,blue] (0,0) -- (vec3) node [below] {$\hat{\mathbf{b}}$};
		\draw[->,thick,blue] (0,0) -- (vec4) node [left] {$\hat{\mathbf{u}}_{1}$};
		\draw[->,thick,red] (0,0) -- (vec5) node [left] {$\mathbf{K}$};
		\draw[dashed,thick,black] (0,0) -- (vec6) ;
		
		\draw [->,black,thick,domain=90:110] plot ({1.5*cos(\x)}, {1.5*sin(\x)}) node[above right] {$\theta$};
		\draw [->,black,thick,domain=110:140] plot ({1.0*cos(\x)}, {1.0*sin(\x)});
		\node[above] at (-0.75,0.75) {$\theta_m$};
		\draw [->,black,thick,domain=180:200] plot ({1.5*cos(\x)}, {1.5*sin(\x)});
		\node[below] at (-1.65,0.0) {$\theta$};
		
		\draw (0,.3)-|(.3,0);
		\draw ({0.3*cos(200)}, {0.3*sin(200)}) -- ({0.3*cos(200)-0.3*cos(70)}, {0.3*sin(200)+0.3*sin(70)}) -- ({0.3*cos(110)}, {0.3*sin(110)});
		
		\draw[blue,thick] (0,-0.5) circle (0.2cm) node[below right] {$\hat{\mathbf{u}}_{2} = \hat{\mathbf{y}}$};
		\draw[blue,thick] (-0.14,-0.64) -- (0.14,-0.36);
		\draw[blue,thick] (0.14,-0.64) -- (-0.14,-0.36);
		
	\end{tikzpicture} \\
	
	\caption{Bases for the probe beam $(\hat{\mathbf{x}}, \hat{\mathbf{y}}, \hat{\mathbf{g}})$ and magnetic field  $(\hat{\mathbf{u}}_1, \hat{\mathbf{u}}_2, \hat{\mathbf{b}})$. The angle between these two coordinate systems is $\theta$ and the angle between the probe beam's wavevector $\mathbf{K}$ and the plane perpendicular to the magnetic field is $\theta_m$, the mismatch angle. Note that $\theta$ and $\theta_m$ are different angles; however, when one is zero, the other is zero as well \cite{Hall-Chen:beam_model_DBS:2022}.
		% $\mathbf{k}_\perp$ and $\mathbf{w}$.
	} 
	\label{fig:basis}
\end{figure}
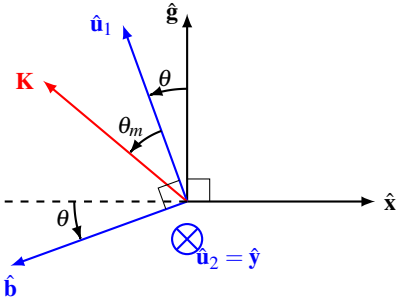

In the rest of paper, we assume that the signal is entirely localised to the cut-off, the location where the probe beam's wavenumber is mimised, and the mismatch attenuates the signal coming from this point. The full model which eschews this assumption, as well as discussing other concerns in interpreting the backscattered signal, like spatial localisation and wavenumber resolution, is presented in earlier work \cite{Hall-Chen:beam_model_DBS:2022}. We briefly summarise the contributions to the detected backscattered spectral density, $p_r$. We have
\begin{equation} \label{eq:A_r2_final_cleaned}
	p_r  (\omega)
	=
	\int 
	F ( l )
	\left<
		\delta n_{e}^2 (t)
	\right>_t
	\widetilde{C} (\omega)
	\ \rmd l ,
\end{equation}
where $l$ is the arc length along the central ray, $\left<\delta n_{e}^2 (t) \right>_t$ is the mean power of the density fluctuations, $\widetilde{C} (\omega)$ is the correlation function, and the rest of the integrand has been abbreviated as $F(l)$. The weighting function $F(l)$ is made of various pieces multiplied together; these pieces relate to the polarisation, mismatch, beam properties, and group velocity \cite{Hall-Chen:beam_model_DBS:2022}. We focus on the mismatch piece
\begin{equation}
	\exp \left[
		- 2 \frac{\theta_{m}^2}{\left(\Delta \theta_{m}\right)^2}
	\right],
\end{equation}
where $\Delta \theta_{m}$ is the mismatch tolerance, given by % Todo: define the correlation function properly.
\begin{equation} \label{eq:delta_theta_m}
	\Delta \theta_{m}
	= 
	\frac{ 1 }{ K } 
	\left[
	\frac{
		\textrm{Im} \left( M_{yy}^{-1} \right)
	}{
		\left[ \textrm{Im} \left( M_{xy}^{-1} \right) \right]^2
		-
		\textrm{Im} \left( M_{xx}^{-1} \right) \textrm{Im} \left( M_{yy}^{-1} \right)
	}
	\right]^{\frac{1}{2}} .
\end{equation} 	
Here we have contributions from the various components of $\bm{M}_w^{-1}$, defined by
\begin{eqnarray}
	\bm{M}_w^{-1}
	=
	\left( \begin{array}{ccc}
		M_{xx}^{-1} & M_{xy}^{-1} & 0 \\
		M_{yx}^{-1} & M_{yy}^{-1} & 0 \\
		0		  & 0		 & 0 \\
	\end{array} \right) \qquad \\
	\qquad = 
	\left(
	\begin{array}{cc}
		&
		\left(
		\begin{array}{cc}
			M_{xx} & M_{xy} \\
			M_{yx} & M_{yy}
		\end{array}
		\right) ^{-1}
		
		\begin{array}{c}
			0 \\
			0 \\
		\end{array} \\
		
		& \begin{array}{cc}
			\ \ \ \ 0 \ \ \ \ & 0~~
		\end{array}

		\begin{array}{c}
			\ \ \ \ \ \ \ \ \ 0
		\end{array}
		
	\end{array} \right) ,
\end{eqnarray}
where
\begin{equation}
	M_{xx} \simeq \Psi_{xx}
	- K \left( \hat{\mathbf{b}} \cdot \nabla \hat{\mathbf{b}} \cdot \hat{\mathbf{g}} \right)
	,
\end{equation}
\begin{equation}
	M_{xy} \simeq \Psi_{xy}
	- K \frac{\left[ \left( \hat{\mathbf{b}} \times \hat{\mathbf{g}} \right)  \cdot \nabla \hat{\mathbf{b}} \cdot \hat{\mathbf{g}} \right]}{\left| \hat{\mathbf{b}} \times \hat{\mathbf{g}} \right|}
	,
\end{equation}
and
\begin{equation}
	M_{yy} = \Psi_{yy}.
\end{equation}
Here $\hat{\mathbf{g}}$ is the unit vector of the group velocity,
\begin{equation}
	\mathbf{g} = \frac{\rmd \mathbf{q}}{\rmd \tau},
\end{equation}
where $\mathbf{q}$ is the position of the central ray and $\tau$ is a coordinate along that ray. The beam's widths and curvatures are related to the real and imaginary parts of $\bm{\Psi}_w$ as given in previous work \cite{Hall-Chen:beam_model_DBS:2022}, and the subscripts indicate their directions, as show in Figure \ref{fig:basis}.

To determine the mismatch piece, we use beam tracing. In this formulation, the electric field is given by\cite{Hall-Chen:beam_model_DBS:2022}
\begin{align} \label{eq:beam_field_final}
	\mathbf{E}_{b}
	&=
	A_{ant} \exp(\rmi \phi_G + \rmi \phi_P) \left[ \frac{\det (\textrm{Im} [\bm{\Psi}_w])  }{\det (\textrm{Im} [\bm{\Psi}_{w,ant} ])} \right]^{\frac{1}{4}}  \sqrt{\frac{g_{{ant}}}{g}} \nonumber  \\
	&\times
	\hat{\mathbf{e}} \exp \left(\rmi s + \rmi \mathbf{K}_w \cdot \mathbf{w} + \frac{\rmi}{2} \mathbf{w} \cdot \bm{\Psi}_w \cdot \mathbf{w}  \right)
	.
\end{align}
Here $\phi_G$ is the Gouy phase, $\phi_P$ is a phase associated with the polarisation, the subscript $_{ant}$ denotes the properties at the launch antenna, and $A_{ant}$ is the initial amplitude. The group velocity, position, wavevector, beam width, and beam curvature are evolved by finding derivatives of the dispersion relation, which we solve with our beam-tracing code, Scotty \cite{Hall-Chen:beam_model_DBS:2022}. This code requires the plasma equilibrium and probe beam's initial conditions as input. The direction of launch, and thus the initial $\mathbf{K} / K$, is given by the poloidal and toroidal launch angles, $\varphi_p$ and $\varphi_t$, respectively. In cylindrical coordinates $(R,\zeta,Z)$, we have
\begin{align}
	K_{R,ant}     &= - \frac{\Omega}{c}         \cos \varphi_t \cos \varphi_p , \\
	K_{\zeta,ant} &= - \frac{\Omega}{c} R_{ant} \sin \varphi_t \cos \varphi_p , \\
	K_{Z,ant}     &= - \frac{\Omega}{c}       				   \sin \varphi_p .
\end{align}
Here $\Omega$ is the beam's angular frequency and $c$ is the speed of light.
	
To gain some physical intuition about the mismatch tolerance, equation (\ref{eq:delta_theta_m}), we consider a circular Gaussian beam propagating through vacuum, and find \cite{Hall-Chen:beam_model_DBS:2022} that equation (\ref{eq:delta_theta_m}) reduces to
\begin{equation} \label{eq:delta_theta_m_vacuum}
	\Delta \theta_{m, vac} 
	= \frac{ \sqrt{2} }{ K W_{0} } .
\end{equation}
Here the subscript $_0$ refers to the beam waist, where the width is minimised. This equation holds as the beam propagates, not just at the waist. The widths and curvatures change in such a way \cite{Hall-Chen:beam_model_DBS:2022} that together, the result is the width at the waist. Note that in earlier work \cite{Hillesheim:DBS_MAST:2015}, the backscattered power was roughly estimated using results from CO$_2$ laser scattering \cite{Slusher:scattering:1980}; the mismatch tolerance is
\begin{equation} \label{eq:delta_theta_m_old}
	\Delta \theta_{m, est} 
	= \frac{ 1 }{ K W } .
\end{equation}
Here $W$ is the beam width. This estimate, $\Delta \theta_{m, est}$ is a factor of $\sqrt{2}$ smaller than that derived from the beam model, equation (\ref{eq:delta_theta_m_vacuum}). When the probe beam is misaligned with the magnetic field, the received backscattered signal is non-zero for two physical reasons. First, there is a spread of wavevectors in the incoming probe beam, due to the Gaussian width and curvature. Hence, even if the main incoming wavevector is mismatched, there will be another wavevector in the probe beam that is matched for $180^{\circ}$ backscattering. Secondly, the receiving antenna has a finite spatial extent. As such, exactly $180^{\circ}$ backscattering is not required for a finite signal. The received backscattered signal has two contributions: the spread of wavevector in the Gaussian beam at the point of scattering and the finite spatial extent of the receiving antenna. The $180^{\circ}$ backscattered power only has the second contribution. Equation (\ref{eq:delta_theta_m_old}) is suitable for use as a conservative design guideline, as was the intention in earlier work \cite{Hillesheim:DBS_MAST:2015}. However, quantitatively accounting for mismatch attenuation of the detected signal requires using the full model.

\subsection{Experimental methodology} \label{subsection:mismatch_experiment}
To investigate the effect of mismatch attenuation on the backscattered signal, the following methodology is typically \cite{Hillesheim:DBS_MAST:2015, Damba:mismatch:2021, Damba:DBS_DIII-D:2022, Hall-Chen:beam_model_DBS:2022} used. Shots are repeated with as similar properties as possible: shape, density, plasma current, and heating. As such, the equilibrium properties are expected to be the same (within margins of error), likewise for the statistical properties of turbulence. The DBS is then steered toroidally for each shot. The poloidal launch angle is fixed and the toroidal launch angle varied. Hence, the probe beams reach similar radial locations and backscatter off similar perpendicular turbulent wavenumbers, but at different mismatch angles, as shown in Figure \ref{fig:toroidal_steering}. The effect of mismatch angle on the signal is thereby ostensibly isolated from the various other complicated effects. In DIII-D, the mismatch angle at the cut-off location varies approximately linearly with toroidal launch angle. However, in MAST and MAST-U, the dependence of $\theta_m$ at the cut-off on the toroidal launch angle is non-trivial. Hence, although one optimizes the toroidal launch angle in experiments, one should remember that the physics actually depends on the mismatch angle, which is indirectly controlled by the toroidal launch angle.

\begin{figure*}
\centering	
\includegraphics[width=0.95\textwidth]{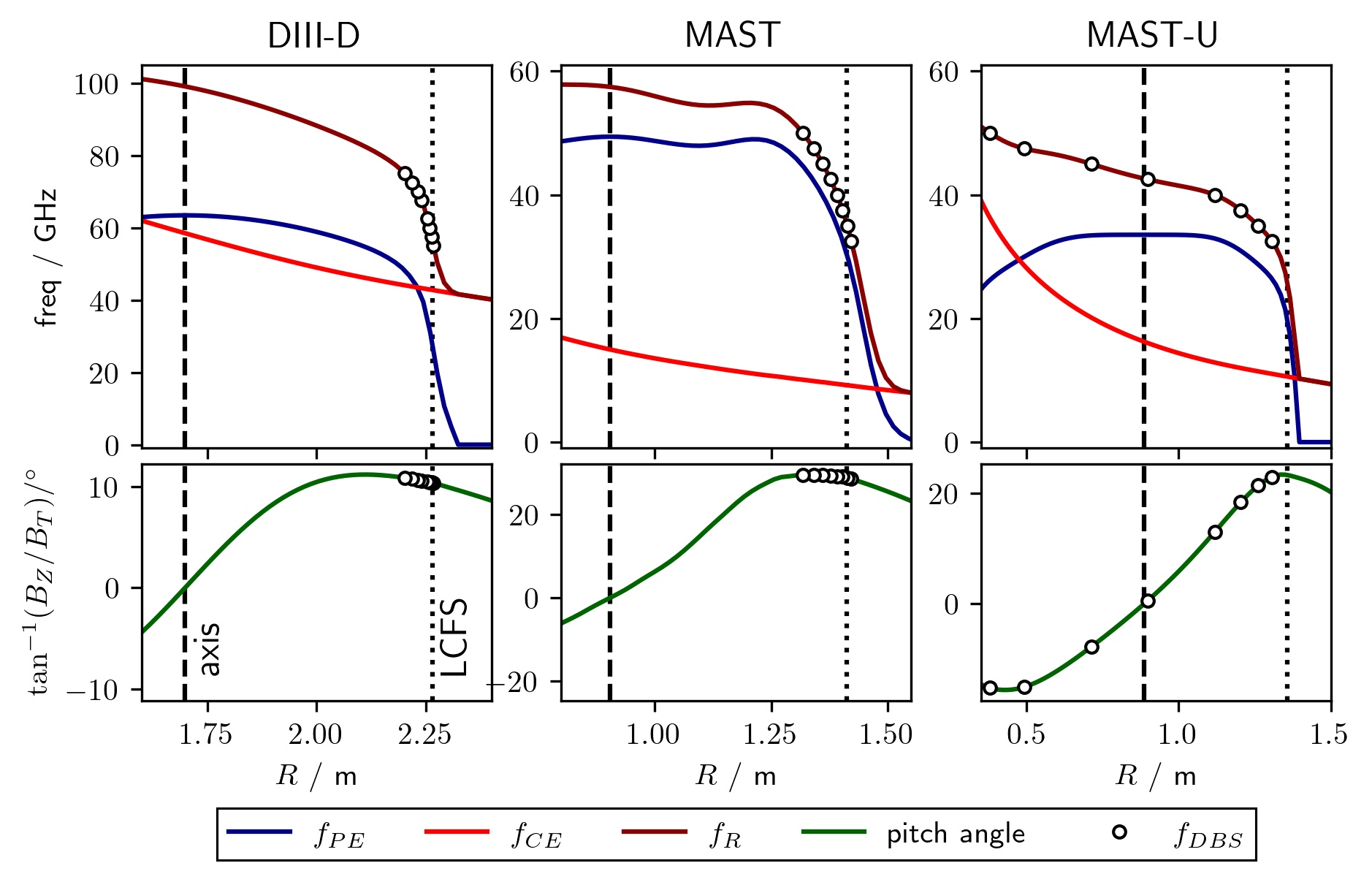}
\caption{Here $f_{PE}$ is the plasma frequency, $f_{CE}$ is the cyclotron frequency, $f_{R}$ is the X-mode cut-off frequency, $\tan^{-1} (B_Z / B_T)$ is the pitch angle, $B_Z$ is a poloidal component of the magnetic field, and $B_T$ the toroidal component. These values of these quantities on the midplane are plotted as a function of radius, $R$. The DBS probe frequencies for DIII-D \cite{Rhodes:DBS:2018} (left), MAST \cite{Hillesheim:DBS_MAST:2015} (middle), and MAST-U (UCLA system) (right) \cite{Rhodes:DBS:2022} are plotted as circles. DIII-D equilibrium data from shot 188839 at 1900 ms, MAST data from shot 29908 at 190 ms, and MAST-U data from shot 45290. All DBS systems were in X-mode for these shots.} 
\label{fig:Equilibria}
\end{figure*}

For each of the three tokamaks (DIII-D, MAST, MAST-U), we study a series of repeated shots with the methodology outlined in the preceding paragraph. Since these tokamaks support a wide range of plasma equilibria, the shots studied in this paper are meant to be illustrative and not definitively representative of all possible plasma configurations. The experimental profiles are shown in Figure (\ref{fig:Equilibria}); more detail is available in other papers on DIII-D \cite{Damba:DBS_DIII-D:2022}, MAST \cite{Hillesheim:DBS_MAST:2015}, and MAST-U \cite{Rhodes:DBS:2022}.

\section{Cross-system study of mismatch attenuation}
Using the methodology described in subsection \ref{subsection:mismatch_experiment}, we evaluate the effectiveness of the beam model in quantitatively account for mismatch in DIII-D, MAST, and MAST-U. The results are then compared with one another.

%[Mismatch angle vs launch angles here]

\subsection{DIII-D} \label{subsection:mismatch_DIII-D}
\begin{figure}
\centering	
\includegraphics[width=\columnwidth]{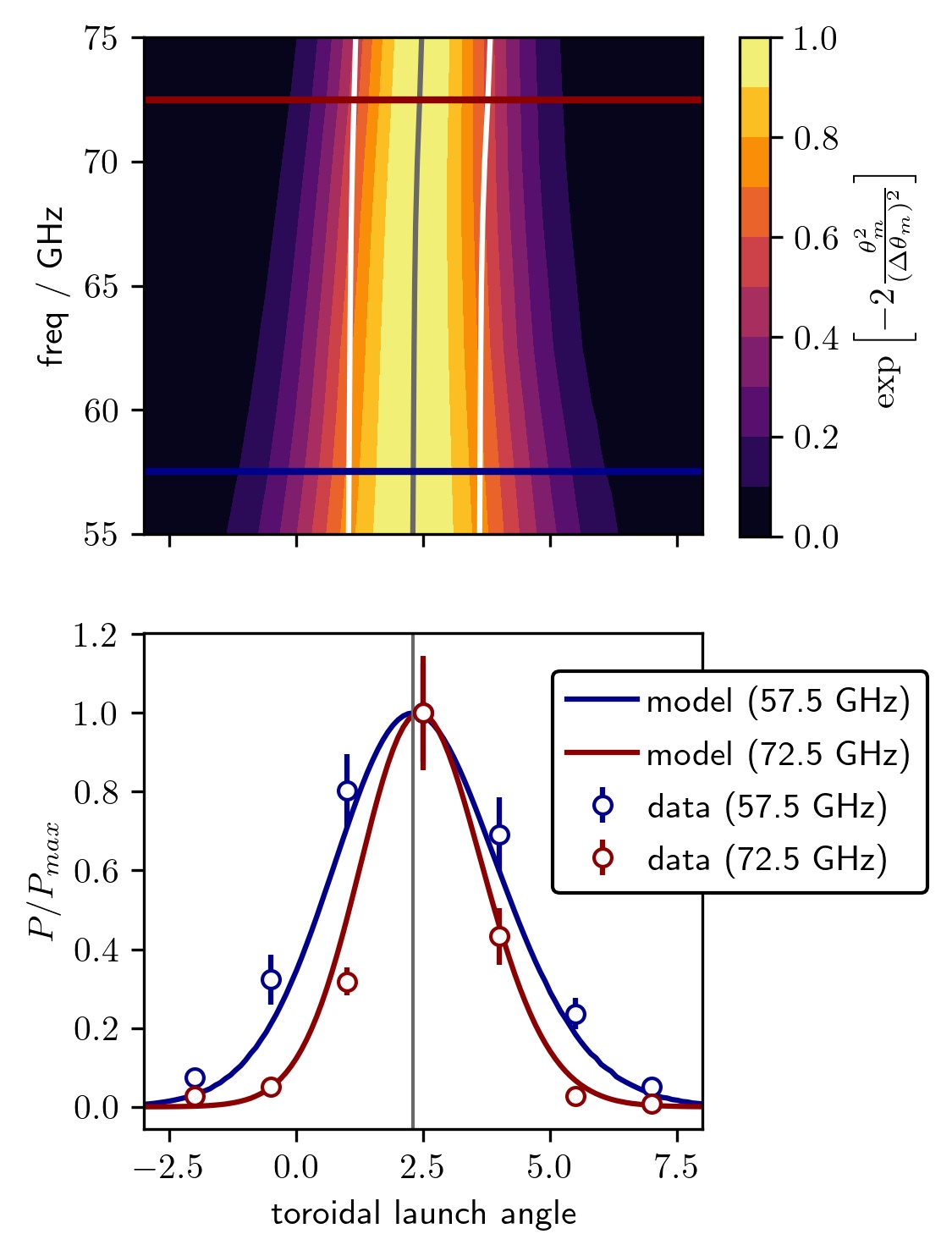}
\caption{The mismatch attenuation piece as a function of toroidal launch angle and frequency (top); both figures share the same x-axis. The gray line gives the $\theta_m = 0$ contour, while the white lines gives the $|\theta_m| = 0.1 \textrm{rad}$ contours (top). The blue and red horizontal lines (top) show the two frequencies where we compare the beam model's results (bottom, solid lines) with experimental data (bottom, circles). The experimental backscattered power, $P$, is normalised to the maximum backscattered power in the toroidal scan, $P_{max}$. Grey vertical line (bottom) shows the angle at which the $57.5$ GHz channel has zero mismatch, the corresponding angle for the $62.5$ GHz channel is about $0.1^{\circ}$ larger. Calculation of error bars discussed in a separate paper \cite{Damba:DBS_DIII-D:2022}. Recall that the beam model is not a fit; the model-based calculation is completely independent of the measured backscattered power data.} 
\label{fig:DIII-D}
\end{figure}

A new DBS system \cite{Rhodes:DBS:2018} was installed on the DIII-D tokamak in 2018. This system is capable of 2D steering of the probe beam; the poloidal and toroidal launch angles can both be independently varied. Toroidal sweeps at fixed poloidal angle were carried out on DIII-D using this DBS system \cite{Rhodes:DBS:2018, Damba:DBS_DIII-D:2022} to investigate the mismatch attenuation. The beam model predicts that mismatch tolerance decreases with increasing wavenumber, Figure \ref{fig:DIII-D} (top). This prediction is validated by experimental data, Figure \ref{fig:DIII-D} (bottom). 
In DBS measurements, one typically attempts to minimise mismatch across all channels \cite{Hillesheim:DBS_MAST:2015, Carralero:DBS_JT60SA:2021}, hopefully achieving reasonable signals across all channels. For the shots studied on DIII-D, the toroidal launch angles to achieve zero mismatch for all channels is very similarly, $2.4\pm 0.1^{\circ}$, Figure \ref{fig:DIII-D}. Hence, it is possible to achieve near-zero mismatch on all channels with an appropriate launch angle \cite{Damba:DBS_DIII-D:2022}. This is likely a consequence of the magnetic pitch angle not varying significantly over the different radial locations that the various channels reach. However, the equilibrium properties change with time; there may be good matching at certain times during the shot, but accounting for mismatch becomes important if one is interested in various times in a shot where the equilibrium properties are sufficiently different. 

\begin{figure*}
\centering	
\includegraphics[width=0.9\textwidth]{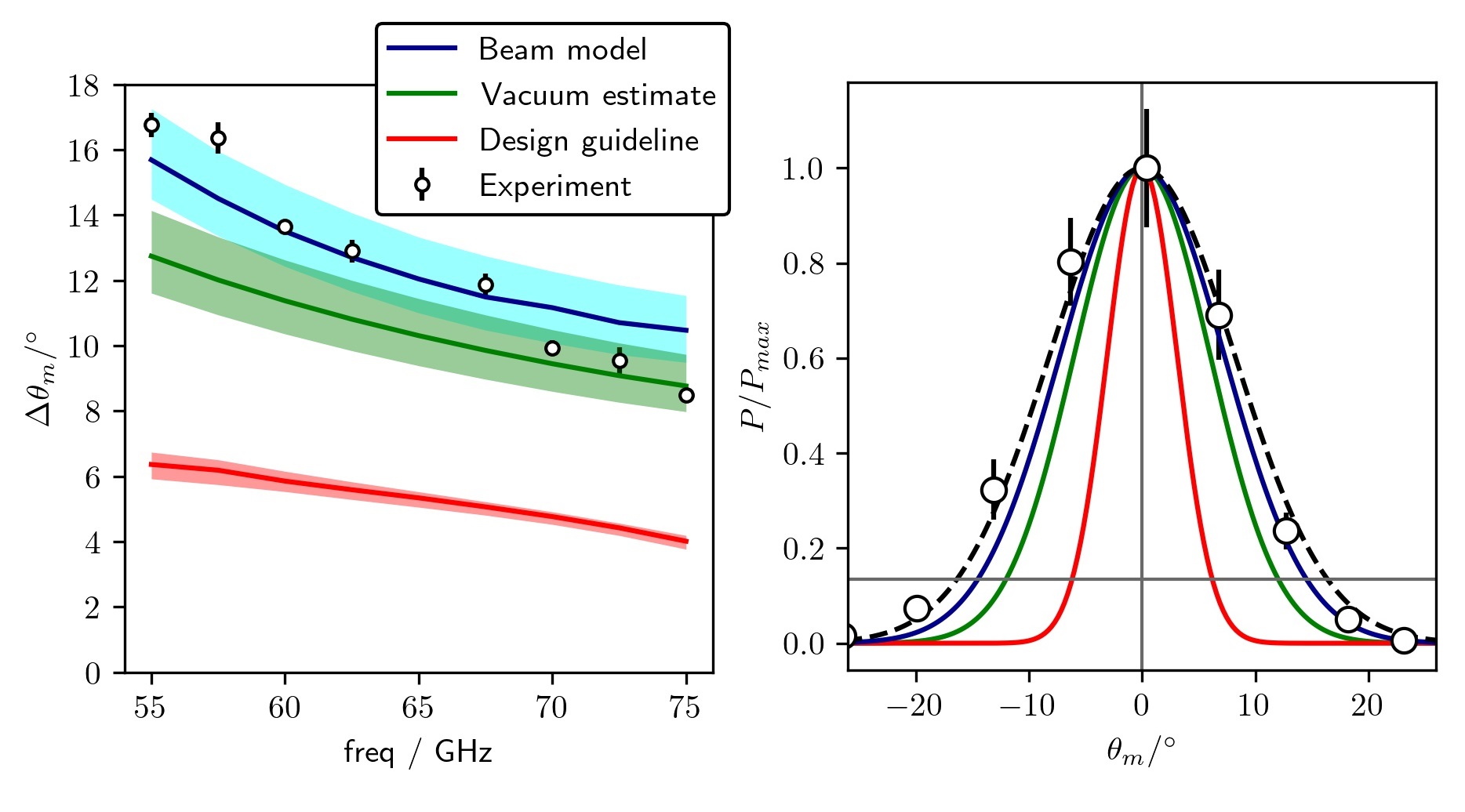}
\caption{The mismatch tolerance in DIII-D as a function of frequency at 180ms (left) and comparison with data for the $57.5$GHz channel (right). Both figures share the same legend. The dotted line (right) indicates the best-fit Gaussian from empirical curve fitting, which is then used for the experimentally-determined mismatch tolerances (left). The experimental backscattered power, $P$, is normalised to the maximum backscattered power in the toroidal scan, $P_max$. The error bars were calculated from the covariance matrices of the curve fitting. The mismatch tolerance is calculated in three different ways: with the full beam model (blue line), equation (\ref{eq:delta_theta_m}), with the plasma wavenumber but beam properties as estimated by vacuum propagation (green line), equation (\ref{eq:delta_theta_m_vacuum}), and with earlier estimates for conservative design guidelines \cite{Hillesheim:DBS_MAST:2015} (red line), equation (\ref{eq:delta_theta_m_old}). The launch widths were varied by $10\%$, shown by the shaded regions (left). Most of the points do indeed lie within the shade blue region, indicating that the beam model does indeed account for mismatch tolerance over all frequencies shown. A Gaussian fit to experimental data is given by the dotted line (right) and the $1 / e^2$ attenuation is given by the horizontal thin black line (right).} 
\label{fig:DIII-D_tolerance_fit}
\end{figure*}

We proceed to quantitatively assess the beam model's ability to account for mismatch attenuation, $\Delta \theta_m$. We calculated $\Delta \theta_m$ with two different methods. Semi-analytically, by using equilibrium data, beam tracing with the Scotty code, and equation \ref{eq:delta_theta_m}. Empirically, by fitting a Gaussian to the backscattered power as a function of toroidal launch angle (black dotted line,  Figure \ref{fig:DIII-D_tolerance_fit} right). The difference in mismatch tolerance as calculated by Scotty and fitting of experimental data, appears to be quite large, up to $2^{\circ}$, Figure \ref{fig:DIII-D_tolerance_fit} (left, difference between points and blue line). However, as we see from Figure \ref{fig:DIII-D_tolerance_fit} (right), a small difference in mismatch tolerance does not make a significant visual difference in the plots of power versus mismatch angle. We varied the launch widths of every frequency by $\pm 10\%$, and find that this does indeed mostly account for the spread of mismatch tolerance. We also varied the launch beam curvature by $\pm 10\%$, but the effect was much less pronounced. Changes in the smoothing schemes of the experimental equilibria resulted in changes to $\Delta \theta_m$ of around $0.5^{\circ}$. It is likely that equilibrium reconstruction might also play a significant role, but we do not examine this in this paper.

Having shown that the beam model can indeed satisfactorily account for mismatch tolerance on DIII-D, we now evaluate other methods of finding the mismatch tolerance. The beam model requires the probe beam's width and curvature as calculated from a beam tracing code. However, beam tracing codes are not yet widely implemented. On the other hand, ray tracing codes are widely available; they can be used to estimate the mismatch attenuation as follows. The wavenumber in equation (\ref{eq:delta_theta_m_vacuum}) is calculated with ray tracing, the beam widths and curvatures are estimated with vacuum propagation, and then magnetic field's shear and curvature are ignored. For the shots studied, this does indeed provide a rough estimate, Figure \ref{fig:DIII-D_tolerance_fit} (green line). We see that the earlier estimates for conservative design guidelines is indeed stricter than our estimate, Figure \ref{fig:DIII-D_tolerance_fit} (red line). Now that we have a quantitative model for mismatch, this design constraint may be somewhat relaxed.
%Prior to the beam model, the mismatch attenuation was estimated \cite{Hillesheim:DBS_MAST:2015} using earlier work on small-angle scattering \cite{Slusher:scattering:1980}, equation (\ref{eq:delta_theta_m_old}); in this formulation, the fit is less than satisfactory, Figure \ref{fig:DIII-D_tolerance_fit} (red lines). The main reason for this is subtle. When the probe beam is misaligned with the magnetic field, the backscattered signal is non-zero for two physical reasons. First, there is a spread of wavevectors in the incoming probe beam, due to the Gaussian width and curvature. Hence, even if the main incoming wavevector is mismatched, there will be another wavevector in the probe beam that is matched for $180^{\circ}$ backscattering. Secondly, the receiving antenna has a finite spatial extent. As such, exactly $180^{\circ}$ backscattering is not required for a finite signal. The reciprocity theorem accounts for both of these contributions to mismatch tolerance, while previous work \cite{Hillesheim:DBS_MAST:2015} does not. 
	
\subsection{MAST} \label{subsection:mismatch_MAST}
\begin{figure}
\centering	
\includegraphics[width=0.95\columnwidth]{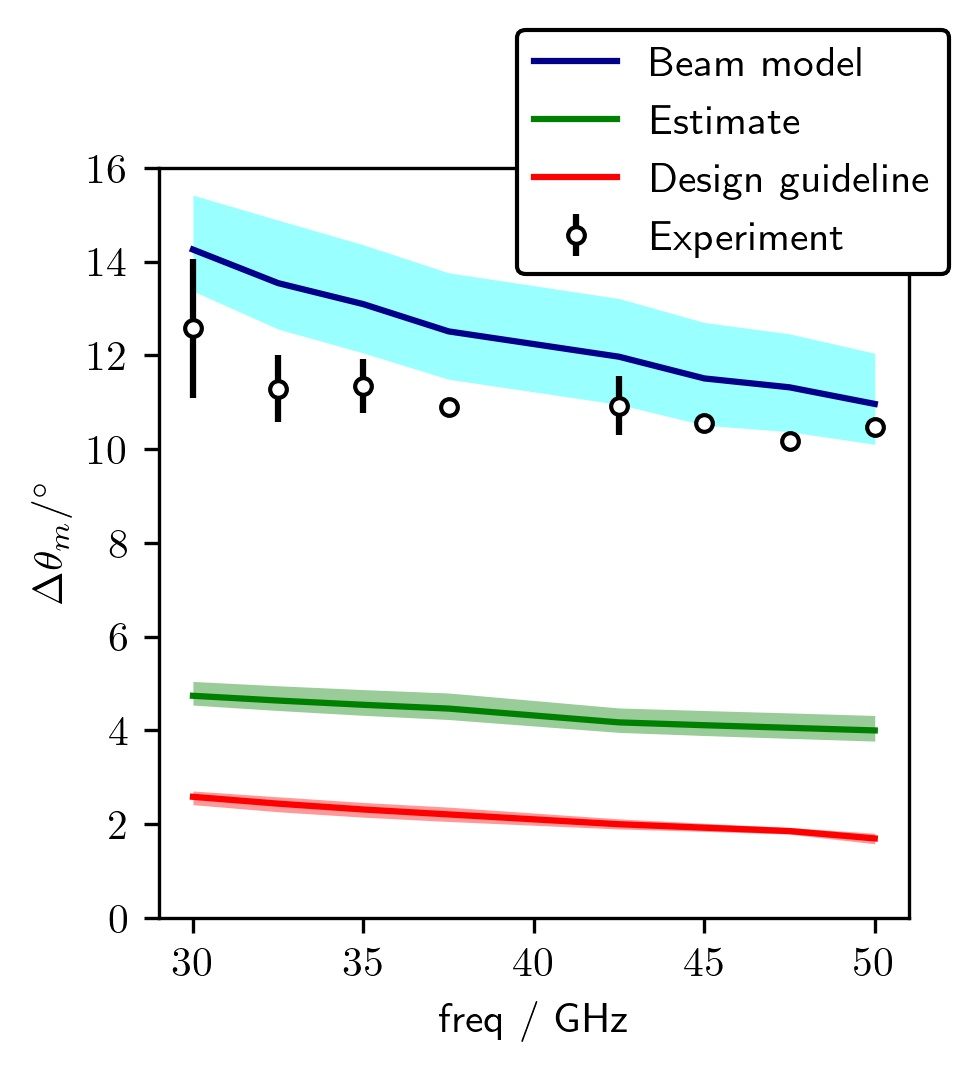}
\caption{The mismatch tolerance in MAST as a function of frequency at $190$ ms. The mismatch tolerance is calculated in four different ways: by curve fitting a Gaussian to experimental data (points), with the full beam model (blue line), with the plasma wavenumber but beam properties as estimated by vacuum propagation (green line), and with earlier estimates for conservative design guidelines \cite{Hillesheim:DBS_MAST:2015, Slusher:scattering:1980} (red line). The launch widths were varied by $10\%$, shown by the shaded regions (left). A systematic error of $3^\circ$ in the steering mirror's rotation angle was used for calculating the experimental points, similar to previous work studying mismatch in MAST \cite{Hillesheim:DBS_MAST:2015}. The vacuum estimate is much poorer for MAST plasmas than DIII-D plasmas (Figure \ref{fig:DIII-D_tolerance_fit}, left).} 
\label{fig:MAST_tolerance}
\end{figure}

The beam model was previously applied to MAST plasmas \cite{Hall-Chen:beam_model_DBS:2022} for a single frequency, 55 GHz, and at a single time, 190 ms. In this work, we analyse the mismatch tolerance of all eight Q-band channels at 190ms, Figure \ref{fig:MAST_tolerance}. As with DIII-D, the mismatch tolerance in MAST is calculated in four different ways: by curve fitting a Gaussian to experimental data, with the full beam model, equation (\ref{eq:delta_theta_m}), with the plasma wavenumber but beam properties as estimated by vacuum propagation, equation (\ref{eq:delta_theta_m_vacuum}), and with pre-beam-model estimate, equation (\ref{eq:delta_theta_m_old}) . The launch beam widths in Scotty were varied by $10\%$ to account for possible errors in their measurement. A systematic error of $3^\circ$ in the steering mirror's rotation angle was used for calculating the experimental points, similar to previous work studying mismatch in MAST \cite{Hillesheim:DBS_MAST:2015}.

There are a few noteworthy points. First, that while Scotty's agreement with experimental data in MAST is decent, Figure \ref{fig:MAST_tolerance}, it is noticeably worse than that of DIII-D, Figure \ref{fig:DIII-D_tolerance_fit} (right). This is possibly due to there being fewer points in the toroidal scan, leading to larger errors. Moreover, the wavenumber at the cut-off varies more strongly with toroidal launch angle for the MAST shots studied than DIII-D, which means the other pieces of the instrumentation weighting function could play a part. Secondly, vacuum propagation poorly estimates the beam's properties and thus the mismatch tolerance, Figure \ref{fig:MAST_tolerance} (green line). For reasons that are not entirely clear, the evolution of the beam is significantly more complicated in MAST than DIII-D. In such a situation, using beam tracing is vital.

\subsection{MAST-U} \label{subsection:mismatch_MAST-U}
There are two DBS systems installed on MAST-U, one from the University of California Los Angeles (UCLA) \cite{Storment:DBS:2021, Rhodes:DBS:2022} and the other from the Southwestern Institute of Physics (SWIP) \cite{Shi:DBS:2019, Wen:DBS:2021}; detailed information about these two systems are available in the cited references. The UCLA system operates in the Q-band and was already installed during MAST-U's first campaign. The SWIP system has two bands, Q and V, and will be installed in time for the second campaign, which has yet to occur at the time of writing. Moreover, its exact quasioptical system might still be modified before installation. As such, the SWIP DBS is not evaluated in this work. 

\begin{figure}
	\centering	
	\includegraphics[width=0.95\columnwidth]{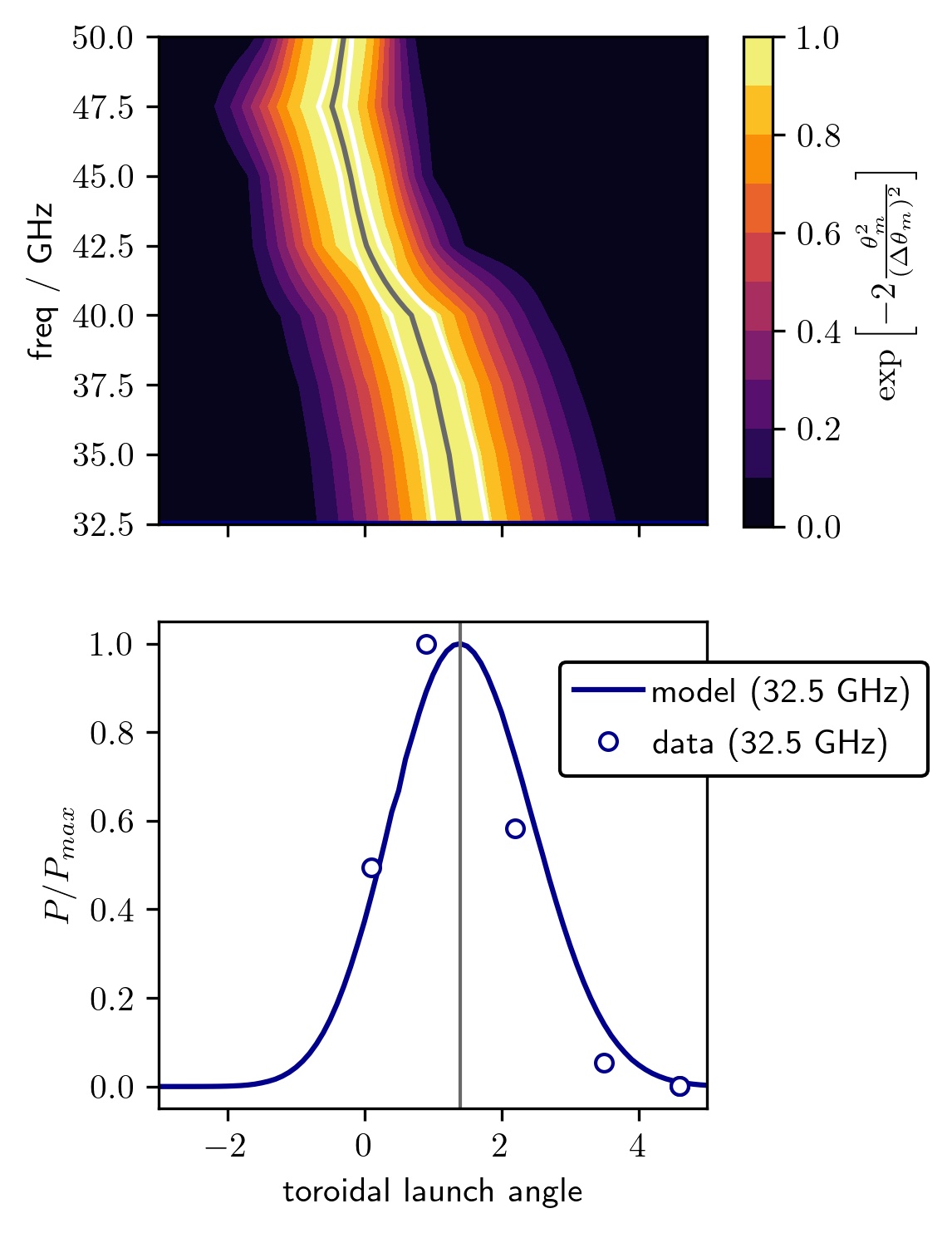}
	\caption{Preliminary results from MAST-U. The mismatch attenuation piece as a function of toroidal launch angle and frequency (top); both figures share the same x-axis. The gray line gives the $\theta_m = 0$ contour, while the white lines gives the $|\theta_m| = 0.1 \textrm{rad}$ contours (top). The experimental backscattered power, $P$, is normalised to the maximum backscattered power in the toroidal scan, $P_{max}$. Grey vertical line (bottom) shows the toroidal launch angle at which the $32.5$ GHz channel has zero mismatch at the cut-off. Since the Q-band frequencies reached significantly different locations with different pitch angles, see Figure \ref{fig:Equilibria}, it is difficult to optimise the toroidal angle for low mismatch across all frequencies simultaneously (top). This is in contrast with DIII-D shot 188839, Figure \ref{fig:DIII-D}.} 
	\label{fig:MAST-U}
\end{figure}

Preliminary results from MAST-U indicate decent agreement between the predicted and measured mismatch tolerances at low frequencies, Figure \ref{fig:MAST-U} (bottom). Since the Q-band frequencies reach significantly different locations with different pitch angles, see Figure \ref{fig:Equilibria}, it is difficult to optimise the toroidal angle for low mismatch across all frequencies simultaneously, Figure \ref{fig:MAST-U} (top). Specifically, for these shots, the channels' measurements span the outboard to the inboard, a dramatic range of locations. This is in contrast with DIII-D shot 188839, Figure \ref{fig:DIII-D}, where all the frequencies reach the cut-off at significantly more similar locations. The situation is made more complicated by the probe beam's wavevector changing directions during propagation and not generally being aligned with the group velocity, which makes beam's wavevector's direction relative to the magnetic field at the cut-off non-trivial. 
At higher frequencies, the optimal toroidal launch angle is significantly smaller,  Figure \ref{fig:MAST-U} (top). As such, there was no clear peak of backscattered power when the toroidal launch angle was varied in the experiments. At least one more point at a negative toroidal launch angle would have been required. Hence, we do not present a systematic comparison of theory and experiment for all frequencies, unlike MAST and DIII-D. The strong dependence of optimal launch angle on frequency poses another challenge: it is difficult to simultaneously optimise mismatch across all channels at once without repeating shots, making the beam model's ability to quantitatively account for mismatch attenuation crucial for interpreting the measured data.

%We now briefly discuss our predictions for mismatch attenuation of the SWIP DBS. Since this system is not yet installed, the finer points of its quasioptical system has yet to be finalised. Consequently, the initial beam conditions has to be estimated and we do not present any Scotty simulations of this system in our paper. That  being said, we expect mismatch attenuation of the SWIP DBS system to be more complicated than UCLA's; the SWIP DBS frequencies span both the Q and V bands, the UCLA DBS's current implementation, which covers only one band. 

\section{Optimizing mismatch tolerance} \label{section:optimisation}
\begin{figure}
	\centering	
	\includegraphics[width=0.95\columnwidth]{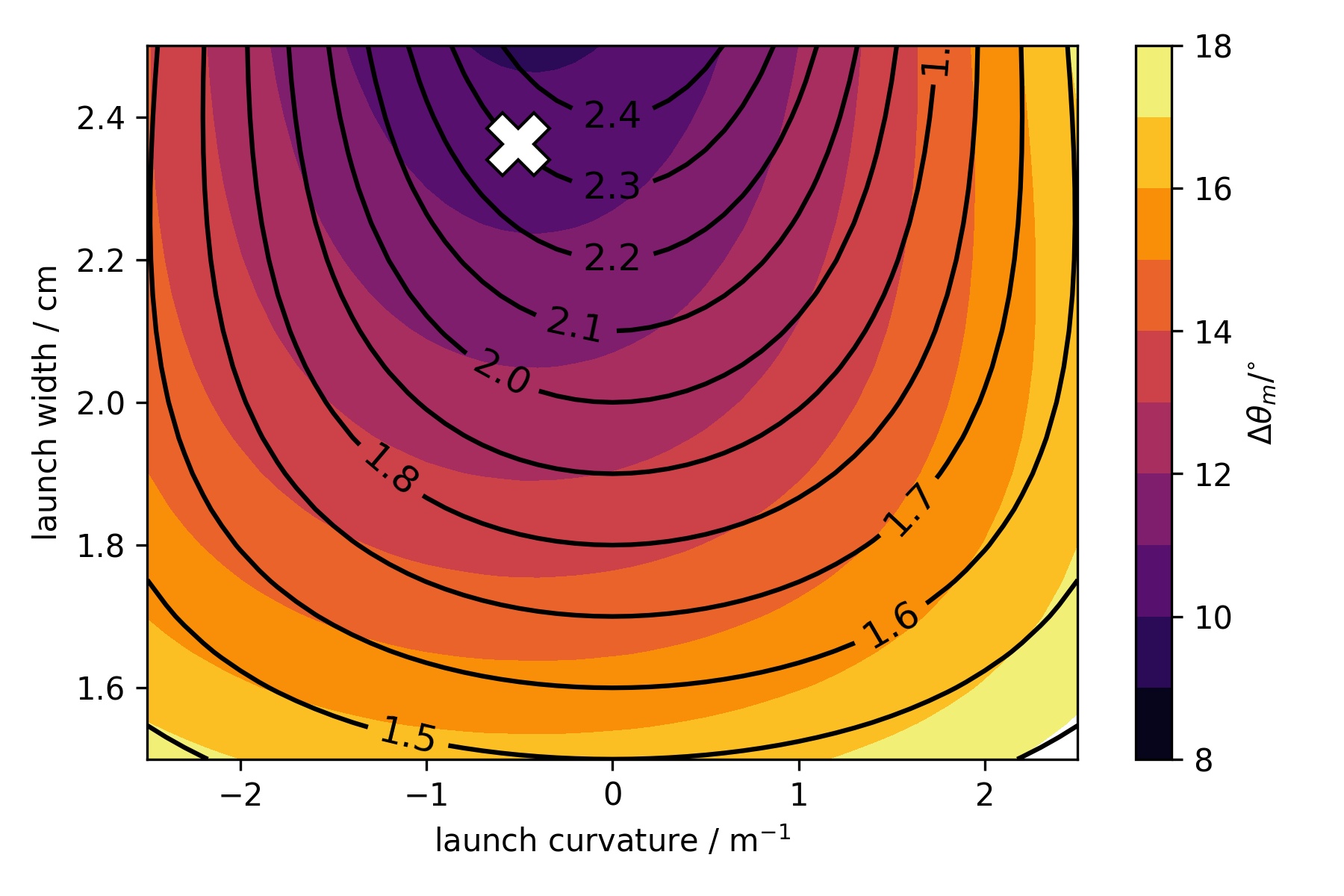}
	\caption{The mismatch tolerance as a function of the width and curvature of the Gaussian probe beam at launch; 75GHz channel on DIII-D. The white cross indicates the beam properties at launch for the current system. The black contours show the size of the beam waist in vacuum, labels are in cm. The relative shift of the black and filled contours are a result of the effects of the plasma on the probe beam. Note that in our simulations, the beam is launched from the steering mirror. The beam's initial properties are specified at these locations, and not at plasma entry.} 
	\label{fig:optimising_tolerance}
\end{figure}

Now that we have established the beam model of DBS as a reliable quantitative method of determining the mismatch attenuation, we can leverage its insights to help us design new DBS systems with more tolerance to mismatch. This is complicated because such optimisation depends on the tokamak and what one is trying to measure. We present preliminary simulations for optimisation in this paper. We consider the DBS on DIII-D and try to see if we can increase the mismatch tolerance by adjusting the launch beam's width and curvature, Figure \ref{fig:optimising_tolerance}. The plasma's effect on the beam's width and curvature plays a significant role in this optimisation, as seen from the difference between the black and filled contours. Note that in our simulations, the beam is launched from the steering optics: the steering mirror for DIII-D and the movable lens for MAST-U. The beam's initial properties are specified at these locations, and not at plasma entry.

Our simulation results show that the mismatch tolerance can be significantly increased. For the particular case studied, this is achieved by having a smaller width as well as a low curvature at launch. It is possible that the parameter space accessible by quasioptical systems is limited, but such analysis is beyond the scope of this paper. Nonetheless, such analysis is likely important when designing new DBS systems, especially when pitch angle at the cut-off varies significantly across the probe frequencies.

\section{Discussion} \label{section:discussion}
The beam model is capable of calculating various contributions to the instrumentation weighting function, $F(l)$. In this work, we evaluate one piece of this function: mismatch attenuation and the associated mismatch tolerance. We see that this approach appears to be more successful at explaining the toroidal scan results from DIII-D than MAST. It is possible that the other pieces of the weighting function \cite{Hall-Chen:beam_model_DBS:2022} play a significant role in MAST's toroidal scan's results. This will be evaluated in a future paper.

Even though using beam tracing and the reciprocity theorem allows one to quantitatively calculate the mismatch attenuation, attempting to get good matching in experiments via prudent toroidal steering is still important for several reasons. First, the signal is larger, which makes it easier to analyse the data. Secondly, as we have seen in this paper, calculating the attenuation has some uncertainty involved, and thus having to correct for it might increase the uncertainty in the measured backscattered power.

\section{Conclusions} \label{section:conclusion}
In this paper, we have shown that the mismatch attenuation in conventional and spherical tokamaks can be quantitatively predicted by using beam tracing and reciprocity theorem. These predictions are validated with experiments on DIII-D, MAST, and MAST-U. Furthermore, we show that the beam model gives significantly better predictions of mismatch tolerance than previous estimates. 

Being able to account for mismatch attenuation increases the versatility of DBS. However, preliminary analysis of MAST-U shows that attempting to get good matching over all channels simultaneously might prove challenging in certain cases, namely, when the pitch angle at the cut-off is significantly different for the various DBS channels. To address this issue, future DBS systems could be designed to have higher mismatch tolerances.

\begin{acknowledgments}
This work was partly supported by the U.S. Department of Energy under contract numbers DE-AC02-09CH11466, DE-SC0019005, DE-SC0019352, DE-SC0020649, and DE-SC0019007. The United States Government retains a non-exclusive, paid-up, irrevocable, world-wide license to publish or reproduce the published form of this manuscript, or allow others to do so, for United States Government purposes. This work has been part-funded by the EPSRC Energy Programme [grant numbers EP/W006839/1 and EP/R034737/1].  To obtain further information on the data and models underlying this paper please contact PublicationsManager@ukaea.uk. \\
\end{acknowledgments}

\appendix

%\section{Appendixes}
%Stuff.

%\nocite{*}
\bibliography{aipsamp}% Produces the bibliography via BibTeX.

\end{document}